\documentclass[
]{ceurart}

\usepackage{listings}
\usepackage{svg}
\begin{document}

\copyrightyear{2025}
\copyrightclause{Copyright for this paper by its authors.
  Use permitted under Creative Commons License Attribution 4.0
  International (CC BY 4.0).}

\conference{CLEF 2025 Working Notes, 9 -- 12 September 2025, Madrid, Spain}

\title{AIRwaves at CheckThat! 2025: Retrieving Scientific Sources for Implicit Claims on Social Media with Dual Encoders and Neural Re-Ranking}

\title[mode=sub]{Notebook for the CheckThat! Lab at CLEF 2025}

\author[1]{Cem Ashbaugh}[%
email=e12433756@student.tuwien.ac.at
]
\fnmark[1]
\cormark[1]
\address[1]{TU Wien, Karlsplatz 13 A-1040 Wien, Austria}

\author[1]{Leon	Baumgärtner}[%
email=e12434736@student.tuwien.ac.at
]
\fnmark[1]

\author[1]{Tim Greß}[%
email=e12412672@student.tuwien.ac.at
]
\fnmark[1]

\author[1]{Nikita Sidorov}[%
email=e62106043@student.tuwien.ac.at
]
\fnmark[0]

\author[1]{Daniel Werner}[%
email=e12429667@student.tuwien.ac.at
]
\fnmark[1]

\cortext[1]{Corresponding author.}
\fntext[1]{These authors contributed equally.}

\begin{abstract}
Linking implicit scientific claims made on social media to their original publications is crucial for evidence-based fact-checking and scholarly discourse, yet it is hindered by lexical sparsity, very short queries, and domain-specific language.  
Team AIRwaves ranked second in Subtask 4b of the CLEF-2025 CheckThat! Lab with an evidence-retrieval approach that markedly outperforms the competition baseline.  
The optimized sparse-retrieval baseline (BM25) achieves MRR@5 $=0.5025$ on the gold label blind test set.
To surpass this baseline, a two-stage retrieval pipeline is introduced: (i) a first stage that uses a dual encoder based on E5-large, fine-tuned using in-batch and mined hard negatives and enhanced through chunked tokenization and rich document metadata; and (ii) a neural re-ranking stage using a SciBERT cross-encoder.  
Replacing purely lexical matching with neural representations lifts performance to MRR@5 $=0.6174$, and the complete pipeline further improves to MRR@5~$=0.6828$.
The findings demonstrate that coupling dense retrieval with neural re-rankers delivers a powerful and efficient solution for tweet-to-study matching and provides a practical blueprint for future evidence-retrieval pipelines.
\end{abstract}


\maketitle
\section{Introduction}

Linking implicit scientific claims on social media back to their original publications is an important capability for evidence‐based fact‐checking, scholarly analysis, and informed public discourse. Social media posts, like tweets, often paraphrase study findings without providing direct identifiers like DOIs or URLs, making automated retrieval a non‐trivial challenge. Hence, in Subtask 4b of the CLEF 2025 CheckThat! Lab \cite{CheckThat:ECIR2025, clef2025-workingnotes}, participants are given a tweet and a fixed pool of candidate papers; the goal is to select the single correct document for each post \cite{clef-checkthat:2025-lncs, clef-checkthat:2025:task4} i.e. tweet‐to‐study matching. Notably, the system developed in the present work ranked second in Subtask 4b of the CLEF 2025 CheckThat! Lab (see the 
\href{https://codalab.lisn.upsaclay.fr/competitions/22359#results}{official leaderboard}). All relevant source code and data can be found on the 
\href{https://github.com/CemAshbaugh/CheckThat-Lab-at-CLEF-2025}{GitHub repository}. Hereinafter, tweets will be referred to as queries and studies as documents.

To tackle the task, a two‐stage IR pipeline is adopted: first, a candidate list is generated by combining BM25 and dual‐encoder retrieval; second, neural re‐ranking algorithms are applied to refine the top‐k results. The experiments on the train, development, and test splits demonstrate that this architecture substantially improves mean reciprocal rank (MRR) and recall over baseline retrievals.
There are three noteworthy challenges inherent to the task at hand:
\begin{itemize}
    \item Lexical sparsity: Tweets omit formal identifiers and often paraphrase study findings, creating a lexical gap between query and document.
    \item Brevity of queries: Short, informal posts limit context for matching against comparatively long abstracts.
    \item Domain specificity: Scientific terminology and varied writing styles demand models that understand technical language in both social media and scholarly text.
\end{itemize}

To guide the exploration of effective retrieval strategies for the task, the following research questions are formulated:

\begin{enumerate}
  \item[\textbf{RQ1:}] Which document metadata is most useful to create embeddings that optimize retrieval performance for the task at hand?
  \item[\textbf{RQ2:}] Does incorporating BM25-mined hard negative training examples for training improve query-to-document retrieval performance?
  \item[\textbf{RQ3:}] To which degree does re-ranking retrieved results using cross-encoders improve the retrieval performance?
\end{enumerate}

The remainder of this paper is structured as follows. Section \ref{sec:related_work} provides an overview of prior approaches: from classical lexical ranking (BM25) and dual encoder retrieval (with in-batch and hard-negative sampling) to neural re-ranking methods (cross-encoders like SciBERT/MedBERT) and multi-stage pipelines that combine sparse and dense retrieval for improved semantic matching. Section \ref{sec:dataset} describes the data at hand. Section \ref{sec:retrieval_pipeline} introduces the two-stage pipeline, combining sparse/dense retrieval with neural re-ranking. Section \ref{sec:reranking} introduces neural re-ranking models and training strategies. Section \ref{sec:results} reports experiments on train, development, and test splits. Finally, Section \ref{sec:conclusion} concludes and outlines future work.

\section{Related Work}
\label{sec:related_work}
As the specific task of post-to-publication matching is not yet well-established in the literature, related areas of research can be considered. In a study with a similar dataset from the IRMiDis track at Forum for Information Retrieval Evaluation 2022, the authors show that BERT significantly outperforms other algorithms in analysing and classifying COVID-19-related tweets, effectively capturing the underlying meaning of the text \cite{bansal2022detecting}. Another relevant approach is presented in the CO-Search system, which was developed to address COVID-19 information retrieval tasks using a multi-stage pipeline combining deep learning-based semantic retrieval (Siamese-BERT) with traditional keyword-based models (BM25), followed by a re-ranking step that integrates question answering and abstractive summarization modules \cite{esteva2021covid}.

The similar task of linking tweets to news articles was introduced by Guo et al. \cite{guo-etal-2013-linking}, who proposed a Weighted Textual Matrix Factorization method for learning representations. Due to fast advancements of deep neural approaches in representation learning, many promising methods have been suggested since. Reimers and Gurevych \cite{reimers-gurevych-2019-sentence} provided foundational work in adapting the BERT architecture to a dual encoder framework well-suited for retrieval tasks, which is the basis of much of the subsequent literature. In the domain of post-to-news-article matching, Danovitch \cite{danovitch2020linking} proposes a custom-made dual encoder to minimize distance between relevant post-article pairs in a joined embedding space while also employing a sequence-length agnostic tokenization technique, which we will adjust to use in our work. Piotrowski et al. \cite{piotrowski2023contrastivenewssocialmedia} present a method differing from the one used in the present work by using separate encoders for posts (Twitter RoBERTa) and articles (RoBERTa), addressing the stylistic mismatch between social media posts with informal language and publications formulated in scientific language.  

Karpukhin et al. \cite{karpukhin-etal-2020-dense} evaluated the influence of using BM25-sampled hard negative documents (i.e. similar to the positive one for a given query) as additional training examples, demonstrating improvements over simple in-batch negatives. In-batch negatives are stated to be a training approach that efficiently uses negative documents from other queries within a batch instead of creating new ones, which is hypothesized to scale well with larger batch sizes \cite{karpukhin-etal-2020-dense}. Gillick et al. \cite{gillick-etal-2019-learning} use a dual encoder approach with in-batch and optional additional hard negative examples, demonstrating high effectiveness even without typical second-stage re-ranking of results retrieved in the first stage. 

Neural re-ranking, in which deep language models reorder an initial candidate list, markedly improves retrieval effectiveness in fact checking and question answering applications \cite{pasin2024seupd}. Mansour et al. \cite{mansour2023verified} proposed a retrieval based re-ranking pipeline that retrieves candidate claims from a corpus of fact-checked statements and reranks them with Sentence-BERT to mitigate the linguistic variability of social media content. 

MacAvenay et al. \cite{macavaney2020sledgez} proposed a two-stage retrieval pipeline with an initial BM25 model that produces candidate-specific papers and a SciBERT-based cross-encoder after for COVID-19 Literature Search. Their study showed that adding a domain-specific transformer re-ranker substantially improved effectiveness, even under zero-shot conditions. In the present work, the same two-stage pattern was followed; however,the retrieval stage is dense rather than sparse.

An even more domain-specific BERT-based language model was introduced by Vasantharajan et al. \cite{medbert}. MedBERT is pretrained on biomedical entity recognition datasets. By leveraging task-specific pretraining, MedBERT was shown to improve downstream biomedical NLP tasks.


\section{Dataset and Experimental Setup}
\label{sec:dataset}
The dataset comprises tweet-study pairs, sourced from the Altmetric corpus and the CORD-19 dataset \cite{wang-etal-2020-cord}. Explicit references such as URLs were removed to simulate real-world scenarios of implicit scientific discourse. Each tweet in the dataset includes informal references to a scientific study, while each scientific paper includes metadata such as the title, abstract, authors, source and journal. The final collection contains 15$\,$699 tweet-study pairs, ensuring a diverse set of claims and associated scientific evidence. For experimentation, the original data split was used, with 12$\,$853, 1$\,$400 and 1$\,$446 datapoints for training, development and test sets, respectively. The document collection contains a total of 7$\,$718 publications. 





\begin{figure}[h]
  \centering
  \def\svgwidth{\linewidth}%
  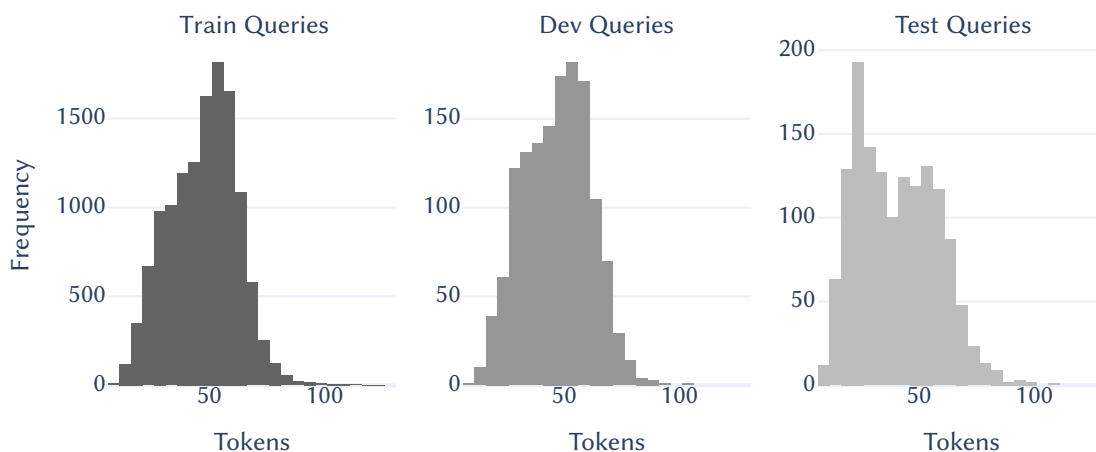
  \caption{Distribution of query token length in train, dev and test datasets.}
  \label{fig:query-length-histograms}
\end{figure}

Figure ~\ref{fig:query-length-histograms} compares the distribution of token lengths across train, dev and test datasets. The histograms visualize the length of the queries in BERT tokens. While the train and dev sets exhibit remarkably similar distributions, with peaks occurring in comparable token length ranges, the test set shows a noticeable shift toward shorter queries. This difference is particularly evident in the higher frequency of queries with 10 to 30 tokens in the test set compared to the other two datasets. The test set's deviation may reflect inherent differences in data collection or curation processes.

The consistent use of BERT tokenization across all datasets rules out tokenization artifacts as a potential cause for this variation, suggesting the differences stem from the underlying data composition itself. The observed length discrepancy between training and test queries should be accounted for during evaluation, as models may exhibit biased performance across different query lengths.

All experiments were conducted on the Lightning AI platform\footnote{Lightning AI. \textit{Lightning AI Platform}. Available at: \url{https://lightning.ai} (accessed May 2025)}, utilizing NVIDIA H100 GPUs (80\,GB memory, 26 CPUs, 1$\,$513 TFLOPs) for resource-intensive tasks and T4 GPUs (16\,GB memory, 8 CPUs, 125 TFLOPs) for standard workloads. For reproducibility, all environment settings such as Torch and Cuda were bound to a random seed.

\section{Methodology}
This section outlines the methodology used to address the scientific claim source retrieval task in the CheckThat! CLEF Challenge 2025 and to answer the research questions guiding the present work. First, the dataset used for the experiments is described. Second, sparse retrieval methods are introduced to establish a baseline for comparison. Third, the proposed two-stage retrieval pipeline is outlined, consisting of the phases neural representation learning and neural re-ranking. Finally, the evaluation metrics applied to assess retrieval performance are presented.

\subsection{Sparse Retrieval: BM25}
\label{sec:sparse_retrieval}

BM25 is a probabilistic Okapi–style ranking function that scores a document \(d\) for a query \(q\) by combining term frequency saturation, document‐length normalization, and inverse‐document frequency \cite{robertson1994some,robertson2009probabilistic}.  Equivalently, it can be written:

\begin{equation}
\mathrm{score}(q,d)
=\sum_{t\in T_d \cap T_q}
\underbrace{\frac{tf_{t,d}}%
     {k_1\bigl((1-b) + b\,\tfrac{dl_d}{avgdl}\bigr) + tf_{t,d}}}_{\substack{\text{TF saturation}\\\text{+ length norm.}}}
\;\times\;
\underbrace{\log\!\Bigl(\frac{|D| - df_t + 0.5}{df_t + 0.5}\Bigr)}_{\text{RSJ IDF}}
\label{eq:bm25}
\end{equation}

where:
\begin{itemize}
  \item \(tf_{t,d}\): frequency of term \(t\) in document \(d\).
  \item \(dl_d\): length of \(d\) in tokens; \(avgdl\): average document length in the collection.
  \item \(df_t\): number of documents containing term \(t\); \(|D|\): total number of documents.
  \item \(k_1 > 0\): term‐frequency saturation parameter (larger \(k_1\) → slower saturation).
  \item \(b \in [0,1]\): length‐normalization parameter (\(b=0\) disables, \(b=1\) full normalization).
\end{itemize}

Conceptually, BM25 refines classic TF–IDF by applying a probabilistic RSJ‐style IDF weight and by normalizing TF according to document length and saturation parameters \cite{robertson2009probabilistic}.  Equation~\ref{eq:bm25} fully specifies the scoring function used in this work.

\paragraph{Preprocessing and indexing.}
Titles and abstracts are concatenated and processed with spaCy, disabling the parser and NER to reduce memory consumption \cite{honnibal2020spacy}.
Tokens are lemmatized, lower-cased, filtered to retain only alphabetic or numeric items, and stop-words are removed using the 547-term list bundled with en\_core\_web\_sm\footnote{\url{https://spacy.io/models/en}}. 
The resulting sequences are indexed with rank-bm25 v0.2.2’s\footnote{\url{https://pypi.org/project/rank-bm25/}} BM25 Okapi implementation.  
The default parameters $(k_{1},b)=(1.5,0.75)$ were used for comparability with the off-the-shelf BM25 model.
An in-memory dictionary caches the top-$k$ lists for previously seen queries, eliminating redundant scoring during evaluation.

\paragraph{Query processing and retrieval.}
Tweets in the query set pass through the same preprocessing pipeline.  
The \(k = 10\) highest-scoring papers are retrieved for each tweet and supplied as the candidate pool for downstream re-rankers.  
The identical cut-off (\(k = 10\)) is used when computing all sparse retrieval metrics reported in Section \ref{sec:Sparse Retrieval Evaluation}.

\paragraph{Rationale.}
Recent large-scale studies demonstrate that, despite advances in dense and hybrid retrieval, BM25 remains a strong, reproducible baseline and often provides complementary or superior signals on short-text tasks \cite{thakur2021beir,wu2022robust}.  
Including it, therefore, furnishes a well-understood reference point and facilitates comparability across systems in CLEF CheckThat! 2025 Task 4b.


\subsection{Retrieval-Pipeline}
\label{sec:retrieval_pipeline}
This section outlines the present work's experiments and their implementation specifics regarding the suggested two-stage retrieval pipeline.
\subsubsection{Phase 1: Neural Representation Learning}
\label{sec:dense_retrieval}
In the following, the details of the experiment implementations for stage 1 are presented. These include the base setup, document metadata experiments and hard negative implementations within the neural representation learning approach. This section addresses the methodology of addressing research questions 1 and 2.
\paragraph{Base Setup.}
\label{sec: meth Base setup}
Initially, a base setup was developed that allows fine-tuning models while keeping all other hyperparameters fixed to determine the most promising dual encoder architecture. Additionally, Word2Vec was explored with a different setup, as it is not a dual encoder architecture. The models considered for this trial are shown in Table \ref{tab:dual-encoders-used}. Word2Vec was applied to learn word embeddings from the corpus to evaluate its suitability for the retrieval task. Two settings were tested: training on the raw text without preprocessing and training on a spaCy-preprocessed corpus, where tokenization, lemmatization, lowercasing, and stopword removal were applied \footnote{\url{https://spacy.io/usage/processing-pipelines}}.

For the base models all-MiniLM-L6-v2\footnote{\url{https://huggingface.co/sentence-transformers/all-MiniLM-L6-v2}} (L6), multi-qa-mpnet-base-dot-v1\footnote{\url{https://huggingface.co/sentence-transformers/multi-qa-mpnet-base-dot-v1}} (MPNet), msmarco-bert-base-dot-v5\footnote{\url{https://huggingface.co/sentence-transformers/msmarco-bert-base-dot-v5}} (MSMarco) and intfloat/e5-large-v2\footnote{\url{https://huggingface.co/intfloat/e5-large-v2}} (E5), the pre-trained implementations from the sentence transformers library are applied.

The rationale behind the choice is, for one, evaluating a diverse set of model parameter sizes (see Table \ref{tab:dual-encoders-used}). L6 was chosen to evaluate the capacity of small models as well as fast, initial prototyping to determine a suitable setup. All models are pre-trained for sentence- and paragraph-matching using different datasets. MPNet and MSMarco excel at multi-domain question-answer retrieval. E5 is the most general-purpose embedding model and the largest in terms of trainable parameters \cite{wang2024-e5}.

\begin{table}[h]
\centering
\caption{Models evaluated using the base setup including their abbreviations used throughout this work.}
\label{tab:dual-encoders-used}
\begin{tabular}{lccc}
\toprule
\textbf{Model} &  \textbf{Abbreviation} & \textbf{Parameters} \\
\midrule
word2vec & W2V & Dataset-dependent \\
all-MiniLM-L6-v2 & L6  & 22.7M \\
multi-qa-mpnet-base-dot-v1 & MPNet  & 109M \\
msmarco-bert-base-dot-v5 &  MSMarco  &  109M  \\
e5-large-v2 &  E5  &  335M  \\
\bottomrule
\end{tabular}
\end{table}

The results for the base setup are shown in Table \ref{tab:results_base_setup}. To verify the influence of batch size on the in-batch negative training setting, the batch sizes [8, 16, 32, 64] were explored using E5. Other hyperparameter settings were chosen according to literature and tests using the dev set. The best learning rate was determined to be $7 \times10^{-6}$. Given the relatively small amount of training data, 2 epochs were sufficient for optimal results, after which overfitting sets in. To prevent large destabilizing updates early in fine-tuning, the first 10\% of all training steps were set as warm-up steps, during which the learning rate is linearly increased to the target value. The distance of learnt document and query embeddings in the shared vector space are calculated using cosine similarity. The construction of input examples and a suitable loss function are crucial components to effective retrieval using dual encoders. Training examples are constructed as (query, positive document) pairs. This allows training on in-batch negatives, which has been found to be effective for similar retrieval tasks \cite{gillick-etal-2019-learning}. For each query, the positive documents of the other queries in the batch serve as negative examples. This can be effectively implemented using MultipleNegativesRankingLoss\footnote{\url{https://sbert.net/docs/package_reference/sentence_transformer/losses.html\#multiplenegativesrankingloss}}. Therefore, the loss function follows formula \ref{eq:mnrl},

\begin{equation}
\label{eq:mnrl}
\mathcal{L} = -\frac{1}{N} \sum_{i=1}^{N} \log \left( 
    \frac{
        \exp\left( \mathrm{scale} \cdot \mathrm{sim}(q_i, p_i) \right)
    }{
        \sum_{j=1}^N \exp\left( \mathrm{scale} \cdot \mathrm{sim}(q_i, p_j) \right)
    }
\right)
\end{equation}

where $N$ is the batch size, $\mathrm{sim}()$ denotes the cosine similarity and $\mathrm{scale}$ is a temperature scaling parameter that is set to 20 in this work. $q_i$ and $p_i$ represent a query-positive document pair. The denominator sums over all positives in the batch, so all other positives serve as in-batch negatives for a given query.

For the base setup exploring the effectiveness of the models in Table \ref{tab:dual-encoders-used} (excluding word2vec), the following preprocessing is applied. Query and document texts are normalized, lower-cased and words are separated by single white spaces. To represent document information, title and abstract are chosen initially. Different document metadata are separated using an explicit [SEP] token. Queries are encoded based solely on the social media post's text. The base setup relies on simple truncation for query and document inputs. If either exceeds the model-intrinsic maximum sequence length (typically 512 tokens for transformer-based architectures), all tokens above the limit are cut off. Table \ref{tab:seq_len_stats} shows the statistics of model input sequence lengths for queries as well as the document lengths using different document metadata combinations.

\begin{table}[h]
\centering
\caption{Token length statistics (min, median, max) for documents and queries after normalization, representing actual model input. 'All fields' refers to title, abstract, authors, journal and source.}
\label{tab:seq_len_stats}
\begin{tabular}{lccc}
\toprule
\textbf{Text Type / Split} & \textbf{Min} & \textbf{Median} & \textbf{Max} \\
\midrule
Title + Abstract        & 19  & 370 & 1824 \\
All fields       & 44  & 424 & 1874 \\
Train Queries           &  8  &  48 & 122  \\
Dev Queries             & 10  &  47 & 102  \\
Test Queries            &  6  &  38 & 131  \\
\bottomrule
\end{tabular}
\end{table}

Queries are never truncated, while documents well above the median token length exceed the limit due to long abstracts. This suggests transformer-based models can benefit from more sophisticated tokenization techniques, especially when incorporating additional document metadata. 

\paragraph{Incorporating Document Metadata.}
\label{meth Incorporating Document Metadata}
To address truncation-induced information loss, the method of chunked tokenization is adapted from Danovitch \cite{danovitch2020linking} and combined with the suggested mean- and max-pooling from Lee, Gallagher and Tu \cite{7927440} to combine the chunks meaningfully. The input sequences are divided into chunks of 510 tokens with an overlap of 50. Out of each chunk, one vector is created through mean-pooling. The individual chunks are then combined into a single fixed-size per-document vector using mean- and max pooling. This embedding creation process for documents is applied during training as well as evaluation and follows formula \ref{eq:chunk-stride-pool},
\begin{equation}
\label{eq:chunk-stride-pool}
\mathbf{h}_{d} = \frac{1}{2} \left( \frac{1}{M} \sum_{i=1}^{M} f(c_i) + \max_{i=1,\ldots, M} f(c_i) \right)
\end{equation}
where $M$ is the number of chunks and $f(c_i)$ the encoder output for chunk $c_i$. The left summand computes the mean-pooled embedding, the right the max-pooled embedding. The combination is meant to combine both the ``typical`` and ``exceptional`` content.

Using this setup, the best-performing model architecture identified using the base setup is used to determine the most informative document metadata. To control for the above-mentioned changes in modelling technique, the original Title + Abstract combination for document metadata was evaluated first, before exploring different combinations. Preprocessing applied depends on the metadata field. Multiple authors and sources are split on ``;``, journal contains single entries and requires no splitting. The used fields of a trial are then combined using [SEP] and normalized as before. The results obtained using chunked tokenization are shown in Table \ref{tab:results-doc-metadata}.

\paragraph{Adding Hard Negatives.}
\label{meth Adding Hard Negatives}
Karpukhin et al. \cite{karpukhin-etal-2020-dense} have shown that using one additional hard negative example per query can outperform training on in-batch negatives only. The rationale is to provide the model with examples that are hard to distinguish, forcing it to learn finer semantic patterns. Gillick et al. \cite{gillick-etal-2019-learning} suggest that this approach is particularly suitable for retrieval tasks where only one document is a correct query partner. This approach is adapted to match the task addressed in this work. BM25 is used to mine hard negative candidates per query. Training examples are now constructed as (query, positive document, hard negative document), where hard negative documents are selected from top BM25 hits for the given query, excluding the positive document. These can be passed to MultipleNegativeRankingLoss, leading to a slightly adjusted formula \ref{eq:mnrl_hn} for the loss,

\begin{equation}
\label{eq:mnrl_hn}
\mathcal{L} = -\frac{1}{N} \sum_{i=1}^{N} \log \left( 
    \frac{
        \exp\left( \mathrm{scale} \cdot \mathrm{sim}(q_i, p_i) \right)
    }{
        \sum_{j=1}^{K} \exp\left( \mathrm{scale} \cdot \mathrm{sim}(q_i, c_{i, j}) \right)
    }
\right)
\end{equation}
where $K$ is the number of all in-batch positives plus any hard negatives passed and $c_{i,j}$ is the $j$-th candidate document for the given query, either an in-batch negative or any included hard negative. In this experiment, exactly one hard negative example was added to each (query, positive document) pair.

Finally, a combination of the aforementioned hard negative and chunked tokenization approach was applied. All hyperparameters are kept fixed and the document fields title, abstract, authors, journal and source were used to construct document representations. The results applying these training approaches are displayed in Table \ref{tab:results-hn}.

\subsubsection{Phase 2: Neural Re-Ranking}
\label{sec:reranking}

After the dual encoder stage had produced the initial candidate list of 100 candidate documents per query, the second stage neural re-ranking module was applied to refine the results. 

\paragraph{Cross-Encoder Transformer Rerankers.}
A cross-encoder jointly processes the query and candidate document, allowing full self-attention between the two texts. Three domain-specific BERT models were fine-tuned:

\begin{itemize}
    \item   DistilBERT \cite{sanh2019distilbert}, which is a distilled, lightweight transformer model optimized for computational efficiency.\footnote{https://huggingface.co/distilbert/distilbert-base-uncased}

    \item   SciBERT \cite{beltagy2019scibert}, a domain-specific transformer pretrained exclusively on scientific tests and therefore capturing scientific terminology and context.\footnote{https://huggingface.co/allenai/scibert\_scivocab\_uncased}

    \item   MedBert \cite{medbert}, is a specialized transformer pretrained on biomedical literature and very suitable for capturing biomedical concepts and nuanced vocabulary.\footnote{https://huggingface.co/Charangan/MedBERT} 
\end{itemize}

The three BERT models were trained in three epochs and hyperparameter tuning on the development set was performed with the batch sizes as well as the learning rates. A learning rate of $2\times10^{-5}$ and a batch size of 16 have proven to be the most effective. Models were trained with a data collator with padding, a max-sequence-length of 512 tokens and a binary cross-entropy loss.

\paragraph{Candidate-set size.}  
Different approaches in literature \cite{reranking_depth, Bttcher2006IndexPA} made clear that more candidate documents do not necessarily mean better performance. To examine the impact of this re-ranking depth, the cross-encoders were run with different numbers of candidate documents between 5 and 100. 

\paragraph{Model Capacity.}  
Table~\ref{tab:reranker_param_sizes} summarizes the parameter counts of all re-ranking models investigated. This overview highlights the latency vs.\ capacity trade‐offs inherent in choosing a reranker for scientific‐claim retrieval.

\begin{table}[h]
\centering
\caption{Re-ranker and their parameter sizes.}
\label{tab:reranker_param_sizes}
\begin{tabular}{llr}
\toprule
\textbf{Model}     & \textbf{Parameters} \\
\midrule
DistilBERT         & 66 M                \\
SciBERT            & 110 M               \\
MedBERT            & 110 M               \\
\bottomrule
\end{tabular}
\end{table}

\subsection{Evaluation Metrics and Significance}
\label{Evaluation Metrics}
Retrieval performance is assessed using MRR@1, MRR@5, MRR@10, Recall@5 and Recall@10. MRR@5 is the official evaluation metric for CheckThat! 2025 Task 4b (Scientific Claim Source Retrieval), as it balances emphasis on early relevant hits with resilience to occasional noisy rankings. MRR@k measures the average position of the single relevant document within the top k, but can be disproportionately influenced by a small number of queries whose relevant document ranks exceptionally high. Recall@k complements MRR by measuring the fraction of queries for which the relevant document appears within the top k, ensuring adequate candidate set coverage for downstream re‐rankers.

Let \(Q\) denote the set of queries and, for each \(q\in Q\), let \(r_q\) represent the 1‐based rank of the single relevant document (or \(+\infty\) if not retrieved).  Define the indicator \(\mathbf{1}(r_q \le k)\), equal to 1 if \(r_q \le k\) and 0 otherwise.  The metrics are:

\begin{align}
\mathrm{MRR}@k 
&= \frac{1}{|Q|} \sum_{q\in Q} \frac{1}{r_q}\,{1}(r_q \le k),
\quad k\in\{1,5,10\},
\label{eq:mrrk}\\[1ex]
\mathrm{Recall}@k 
&= \frac{1}{|Q|} \sum_{q\in Q} {1}(r_q \le k),
\quad k\in\{5,10\}.
\label{eq:recallk}
\end{align}

\noindent where:
\begin{itemize}
  \item \(Q\): set of all queries.
  \item \(|Q|\): total number of queries.
  \item \(r_q\): 1‐based rank of the relevant document for query \(q\) (or \(+\infty\) if not returned).
  \item \(k\): cut‐off depth (\(k\in\{1,5,10\}\) for MRR; \(\{5,10\}\) for Recall).
  \item \({1}(r_q \le k)\): indicator function, 1 if the relevant document is in the top-\(k\), 0 otherwise.
\end{itemize}

Higher MRR@k indicates that relevant documents tend to appear closer to the top (with positions beyond k contributing zero), while higher Recall@k indicates that a larger fraction of queries retrieve their relevant document within the top‐k results.  

To evaluate the significance of difference in test set performance relevant to the research questions, the Wilcoxon signed-rank test is applied for MRR@5 and McNemar's test for MRR@1. For research question 1, we compare the E5 predictions from the control trial (title, abstract) with the trial using all metadata fields (title, abstract, authors, journal, source). To address research question 2, the base setup E5 predictions are compared with the E5 trained using additional hard negative examples. For research question 3, we compare the base setup E5 predictions with the same predictions after re-ranking using SciBERT. Note that some violations of the Wilcoxon assumptions, particularly symmetry of the differences of per-query MRR@5 scores, cannot be completely ruled out.

\section{Results}
\label{sec:results}
This section presents the results and analyses of experiments run in stage 1 and stage 2. It provides an overall comparison of the most promising methods applied. Most notably, different neural representation learning approaches, such as hard negative learning regimes and neural re-ranking, are applied.
\subsection{Sparse Retrieval Evaluation}
\label{sec:Sparse Retrieval Evaluation}

Table~\ref{tab:bm25_results} reports the effectiveness of the lexical baseline under two configurations: (i) raw BM25 tokenizing on whitespace only (None) and (ii) the identical ranker coupled with the spaCy lemmatization and stop-word removal pipeline (spaCy).

\paragraph{Impact of preprocessing.}
Across all splits, the spaCy variant dominates the no-preprocessing run on every metric.  
Focusing on the leaderboard metric MRR@5, lemmatization and stop-word removal raise performance by

\begin{itemize}
\item Train: +0.060 absolute (0.5510 \(\rightarrow\) 0.6109) and +0.0762 Recall@5;
\item Dev:   +0.063        (0.5523 \(\rightarrow\) 0.6157) and +0.0736 Recall@5;
\item Test:  +0.071       (0.4314 \(\rightarrow\) 0.5025) and +0.0892 Recall@5.
\end{itemize}
These improvements confirm that normalizing inflectional variants and removing high-frequency function words reduces term–document noise and helps lexical matching.

\paragraph{Rank‐depth trend.}  

MRR increases from ranks @1 to @5, rising from +0.0429 to +0.0503 in the raw variant and from +0.0507 to +0.0640 in the spaCy variant, but then increases by only +0.0044 to +0.0054 in the raw variant and by +0.0062 to +0.0072 in the spaCy variant between ranks @5 and @10, indicating that virtually all useful gains occur within the first five ranks.

Recall increases from +0.0318 to +0.0393 for raw and from +0.0463 to +0.0526 for spaCy between @5 and @10, confirming that depth beyond five documents yields diminishing returns.  
Pre‐processing therefore enhances retrieval exactly at the cut‐off that determines the challenge ranking, while improvements at deeper ranks remain limited.

\begin{table}[h]
\centering
\caption{Retrieval performance of the BM25 baseline with k = 10 on train, dev, and test sets with and without the spaCy preprocessing pipeline.}
\label{tab:bm25_results}
\begin{tabular}{llccccc}
\toprule
\textbf{Dataset} & \textbf{Preprocessing} & \textbf{MRR@1} & \textbf{MRR@5} & \textbf{MRR@10} & \textbf{Recall@5} & \textbf{Recall@10} \\
\midrule
\textbf{Train} & None   & 0.5081 & 0.5510 & 0.5560 & 0.6171 & 0.6542 \\
               & spaCy  & 0.5585 & \textbf{0.6109} & 0.6171 & 0.6933 & 0.7396 \\
\cmidrule{1-7}
\textbf{Dev}   & None   & 0.5057 & 0.5523 & 0.5577 & 0.6250 & 0.6643 \\
               & spaCy  & 0.5650 & \textbf{0.6157} & 0.6221 & 0.6986 & 0.7479 \\
\cmidrule{1-7}
\textbf{Test}  & None   & 0.3811 & 0.4314 & 0.4358 & 0.5090 & 0.5408 \\
               & spaCy  & 0.4385 & \textbf{0.5025} & 0.5097 & 0.5982 & 0.6508 \\
\bottomrule
\end{tabular}
\end{table}

\paragraph{Field-weighted variant.}

Metadata‐aware extensions to BM25 have been shown to offer negligible benefits over the original formulation, with large‐scale reproducibility studies reporting no significant differences in retrieval effectiveness \cite{kamphuis2020}. Consequently, the standard BM25 configuration is adopted as the primary baseline.

\paragraph{Generalization gap.}
Absolute scores drop from Train/Dev to Test under both configurations, indicating a distribution shift between the public splits and the blind evaluation corpus.  
Nevertheless, the relative advantage of the spaCy pipeline is preserved, suggesting that the linguistic normalization it provides is robust to topic drift.

\paragraph{Head-room for dense methods.}
Even with preprocessing, Recall@10 reaches only 0.6508 on Test, leaving substantial room for downstream neural re-rankers or hybrid retrieval to improve coverage.  
The sparse baseline therefore offers a realistic yet reproducible starting point for subsequent components in the pipeline, and the following evaluation sections will demonstrate how more advanced (dense and hybrid) approaches close this performance gap.

\subsection{Retrieval Pipeline}
\label{Results Retrieval Pipeline}
In the following paragraphs, the results of the neural representation learning experiments are presented and evaluated. Subsequently, the re-ranking of the best predictions generated in that step is evaluated. 
\subsubsection{Phase 1: Neural Representation Learning}
\label{Results Phase 1: Neural Representation Learning}

\paragraph{Base setup results.}
\label{Base setup results}
Table~\ref{tab:results_base_setup} presents the retrieval performance of different models when trained using the base setup, including traditional Word2Vec embeddings and different dual encoder models, across the training, development, and test sets. 

For Word2Vec, the performance improved with spaCy preprocessing, as it did for the sparse retrieval model in Section \ref{sec:Sparse Retrieval Evaluation}. It was also tested to use spaCy preprocessing for the transformer-based sentence embedding models. The tests showed, as Haviana et al. \cite{10295797} also discussed, that preprocessing steps such as lemmatization and stopword removal do not improve but rather degrade performance for transformer-based sentence embedding models because the models otherwise are not able to capture the rich contextual information. Therefore, the results in the table show Word2Vec with spaCy preprocessing and the other models without spaCy preprocessing. 

\begin{table}[h]
\centering
\caption{Performance of models trained using the base setup.}
\label{tab:results_base_setup}
\begin{tabular}{lllccccc}
\toprule
\textbf{Model} & \textbf{Fine-Tuned} & \textbf{Dataset} & \textbf{MRR@1} & \textbf{MRR@5} & \textbf{MRR@10} & \textbf{Recall@5} & \textbf{Recall@10} \\
\midrule
\multirow{3}{*}{W2V} 
& \multirow{3}{*}{No} 
& Train & 0.1099 & 0.1516 & 0.1610 & 0.2262 & 0.2967 \\
& & Dev  & 0.1107 & 0.1450 & 0.1551 &  0.2071  & 0.2836 \\
& & Test & 0.0719 & 0.1059 & 0.1128 & 0.1646 & 0.2172 \\
\midrule
\multirow{3}{*}{L6}
& \multirow{3}{*}{Yes} 
& Train & 0.5007 & 0.5729 & 0.5817 & 0.6879 & 0.7537 \\
& & Dev  & 0.5229 & 0.5877 & 0.5961 & 0.6929 & 0.7543 \\
& & Test & 0.4253 & 0.4999 & 0.5076 & 0.6183 & 0.6770 \\
\midrule
\multirow{3}{*}{MPNet}
& \multirow{3}{*}{Yes} 
& Train & 0.5901 & 0.6608 & 0.6680 & 0.7712 & 0.8248 \\
& & Dev & 0.5821 & 0.6470 & 0.6540 & 0.7529 & 0.8043 \\
& & Test & 0.4696 & 0.5480 & 0.5554 & 0.6715 & 0.7268 \\
\midrule
\multirow{3}{*}{MSMarco}
& \multirow{3}{*}{Yes} 
& Train & 0.5327 & 0.6055 & 0.6142 & 0.7223 & 0.7862 \\
& & Dev & 0.5364 & 0.6005 & 0.6074 & 0.7043 & 0.7557 \\
& & Test & 0.4246 & 0.4984 & 0.5069 & 0.6155 & 0.6777 \\
\midrule
\multirow{3}{*}{E5}
& \multirow{3}{*}{No}
& Train & 0.5526 & 0.6221 & 0.6290 & 0.7289 & 0.7805 \\
& & Dev & 0.5829 & 0.6428 & 0.6499 & 0.7364 & 0.7879 \\
& & Test & 0.4447 & 0.5183 & 0.5280 & 0.6349 & 0.7054 \\
\midrule
\multirow{3}{*}{E5}
& \multirow{3}{*}{Yes} 
& Train & 0.6619 & \textbf{0.7369} & 0.7432 & 0.8491 & 0.8952 \\
& & Dev & 0.6436 & \textbf{0.7029} & 0.7104 & 0.7921 & 0.8457 \\
& & Test & 0.5166 & \textbf{0.6018} & 0.6100 & 0.7317 & 0.7946 \\
\bottomrule
\end{tabular}
\end{table}

It is evident that Word2Vec performs significantly worse than the transformer-based models. Even without fine-tuning, E5 outperforms all other models except MPNet.

Without fine-tuning, E5 results in an MRR@5 of 0.5183, while its fine-tuned counterpart results in an MRR@5 of 0.6018. This confirms that, in line with the findings of \citet{rathinasamy2024e5}, fine-tuning the E5 model yields significantly greater retrieval performance. The fine-tuned E5 performs noticeably better than all other models, followed by MPNet. MSMarco and L6 achieve comparable scores despite both being pre-trained on general text sequence matching and the L6 model having much fewer trainable parameters, stressing the efficiency of the latter.

Across models and metrics, there is a small gap in performance between the train and dev set, suggesting mild overfitting. The discrepancy between train and test set, however, is around 0.1 or upwards in all cases. This can be attributed to several possible reasons. One of these is a shift in the test data distribution, which was already hinted at in Figure \ref{fig:rank-distribution}.

For the best-performing model E5, batch size experiment results for the test set are shown in Table \ref{tab:e5_batchsize_test}, where the setup corresponds to the base setup, only varying a single hyperparameter.

\begin{table}[h]
\centering
\caption{Test set retrieval performance for E5 at various batch sizes.}
\label{tab:e5_batchsize_test}
\begin{tabular}{l c c c c c c}
\toprule
\textbf{Model} & \textbf{Batch size} & \textbf{MRR@1} & \textbf{MRR@5} & \textbf{MRR@10} & \textbf{Recall@5} & \textbf{Recall@10} \\
\midrule
E5 & 8  & 0.5104 & 0.5933 & 0.6024 & 0.7206 & 0.7898 \\
E5 & 16 & 0.5090 & 0.5922 & 0.6007 & 0.7213 & 0.7870 \\
E5 & 32 & 0.5187 & \textbf{0.6037} & 0.6120 & 0.7317 & 0.7932 \\
E5 & 64 & 0.5166 & 0.6018 & 0.6100 & 0.7317 & 0.7946 \\
\bottomrule
\end{tabular}
\end{table}
The results suggest a slight upward trend, with all test metrics resulting from a batch size of 64 exceeding batch size 8 metrics, showing some fluctuations with intermediate batch sizes. This confirms the understanding of Karpukhin et al. \cite{karpukhin-etal-2020-dense}, that in-batch negative training settings benefit from larger batch sizes, as it increases the ratio of negative/positive training examples for any given query.

\paragraph{Incorporating Document Metadata.}
\label{Results Incorporating Document Metadata}
Table \ref{tab:results-doc-metadata} shows the test set performance of E5 using chunked tokenization with max- and mean-pooling. 
\begin{table}[ht]
\centering
\caption{E5 test set retrieval scores for different document field combinations using chunked tokenization and max- and mean-pooling. 'All fields' refers to title, abstract, authors, journal and source.}
\label{tab:results-doc-metadata}
\begin{tabular}{lccccc}
\toprule
\textbf{Document Fields} & \textbf{MRR@1} & \textbf{MRR@5} & \textbf{MRR@10} & \textbf{Recall@5} & \textbf{Recall@10} \\
\midrule
title+abstract (control) & 0.5166 & 0.6020 & 0.6102 & 0.7324 & 0.7960 \\
title+abstract+authors   & 0.5256 & \textbf{0.6089} & 0.6173 & 0.7351 & 0.8001 \\
title+abstract+journal   & 0.5194 & 0.6031 & 0.6117 & 0.7310 & 0.7974 \\
title+abstract+source    & 0.5207 & 0.6049 & 0.6127 & 0.7344 & 0.7960 \\
all fields               & 0.5228 & 0.6069 & 0.6152 & 0.7351 & 0.7994 \\
\bottomrule
\end{tabular}
\end{table}

The control trial reveals that applying chunked tokenization and max- and mean pooling to combine chunks yields no substantial change in test set performance measured by MRR@5 in comparison to the base setup E5 (Table \ref{tab:results_base_setup}) while using the same document fields. This finding is somewhat surprising, as we showed in Table \ref{tab:seq_len_stats} that some documents exceed the maximum sequence length, confirming that information is lost using the base setup without chunked tokenization. This might be attributed to only little meaningful abstract information being truncated using the base setup. For instance, since not all documents in the collection are necessarily mentioned in any queries, long documents might simply not be referenced often in our dataset. 

Adding any additional field individually provides an improvement between 0.0069 and 0.0011, with the authors field contributing the most information and journal the least. However, it can be concluded that title and abstract convey the most relevant information to match queries and documents, while author, journal and source information only improve performance marginally compared to the control trial. Finally, using all considered fields together improves test set MRR@5 by 0.0049. The control trial and all-fields trial correctly rank roughly 51.7\% and 52.3\% of positive documents in first place, respectively, as measured by MRR@1. To determine the significance of the test set differences (control vs. all fields) in MRR@1 and MRR@5, McNemar's test and the Wilcoxon signed-rank test are applied, yielding p-values of 0.2430 and 0.0715, respectively. As both are larger than $\alpha=0.05$, we conclude that there is no evidence of a significant difference in retrieval performance when adding further document metadata in addition to title and abstract. 

\paragraph{Adding Hard Negatives.}
\label{Results adding Hard Negatives}
Table \ref{tab:results-hn} shows the retrieval metrics for E5 when incorporating hard negative training examples as well as combining hard negative examples with additional document metadata.
\begin{table}[ht]
\centering
\caption{Retrieval scores for the E5 hard negative implementation and hard negative implementation and all document fields combined across train, dev, and test sets.}
\label{tab:results-hn}
\begin{tabular}{llcccccc}
\toprule
\textbf{Model} & \textbf{Dataset} & \textbf{MRR@1} & \textbf{MRR@5} & \textbf{MRR@10} & \textbf{Recall@5} & \textbf{Recall@10} \\
\midrule
\multirow{3}{*}{E5+HN}
& Train & 0.6992 & 0.7669 & 0.7723 & 0.8685 & 0.9083 \\
& Dev   & 0.6393 & 0.7014 & 0.7071 & 0.7964 & 0.8393 \\
& Test  & 0.5353 & 0.6171 & 0.6256 & 0.7434 & 0.8064 \\
\midrule
\multirow{3}{*}{E5+HN+All Fields}
& Train & 0.7025 & \textbf{0.7698} & 0.7750 & 0.8705 & 0.9089 \\
& Dev   & 0.6443 & \textbf{0.7079} & 0.7127 & 0.8021 & 0.8386 \\
& Test  & 0.5332 & \textbf{0.6174} & 0.6254 & 0.7469 & 0.8064 \\
\bottomrule
\end{tabular}
\end{table}
We found that adding one hard negative document per query generally results in stronger overfitting compared to using only in-batch negatives as in the base approach, as reflected in the relatively larger gap between train and dev scores for both E5+HN and E5+HN+All fields. Despite this, the test score measured by MRR@1 and MRR@5 is improved from 0.5166 and 0.6018 to 0.5353 and 0.6171 when providing one additional hard negative, respectively, when compared to the base setup E5. To test the significance of the test set MRR@1 and MRR@5 performance gains between the base setup E5 and E5+HN, we apply the same tests as before. McNemar's test for MRR@1 and the Wilcoxon signed-rank test for MRR@5 yield p-values of 0.0159 and 0.0012, respectively, falling below $\alpha=0.05$ in both cases. We therefore conclude that applying the hard negative training regime in addition to in-batch negatives significantly improves retrieval performance.

The positive document for a query is ranked roughly 1.2\% more often in the top 5 predictions with E5+HN than in the base setup. By contrast, additionally providing a hard-negative setup with the additional document fields authors, journal and source (row with E5+HN+All Fields) yields comparable results as E5+HN across the board.

\subsubsection{Re-Ranking}
The following paragraphs report the results of the implemented neural re-ranking experiments.

\paragraph{Comparison of Re-ranking Models.}  
To quantify the influence of the cross-encoder on re-ranking quality, we applied three models—DistilBERT, SciBERT, and MedBERT—to the candidate lists generated by the baseline E5 dual encoder. Table \ref{tab:reranker_comparison} reports mean reciprocal rank at one and five (MRR@1, MRR@5) and recall at five (Recall@5) on the official train, development, and test partitions. All hyperparameters, including the batch size of 16 and the top-5 candidate depth, were held constant across runs.

\begin{table}[h]
\centering
\caption{Retrieval performance of rerankers (top-5 candidates, batch size = 16).}
\label{tab:reranker_comparison}
\begin{tabular}{llccc}
\toprule
\textbf{Model} & \textbf{Dataset} & \textbf{MRR@1} & \textbf{MRR@5} & \textbf{Recall@5} \\
\midrule
\multirow{3}{*}{SciBERT}    
& Train & 0.8216 & \textbf{0.8338} & 0.8490 \\
& Dev   & 0.6507 & \textbf{0.7070} & 0.7921 \\
& Test  & 0.6113 & \textbf{0.6607} & 0.7317 \\
\midrule
\multirow{3}{*}{DistilBERT} 
& Train & 0.7615 & 0.7973 & 0.7615 \\
& Dev   & 0.6343 & 0.6954 & 0.6343 \\
& Test  & 0.5705 & 0.6340 & 0.7317 \\
\midrule
\multirow{3}{*}{MedBERT}    
& Train & 0.7954 & 0.8178 & 0.8490 \\
& Dev   & 0.6043 & 0.6747 & 0.7921 \\
& Test  & 0.5823 & 0.6381 & 0.7317 \\
\bottomrule
\end{tabular}
\end{table}

Across all splits, SciBERT achieved the strongest overall performance, reaching MRR@5 scores of 0.8338, 0.7070, and 0.6607 on the train, development, and test sets, respectively, with corresponding Recall@5 values of 0.8490, 0.7921, and 0.7317. DistilBERT yielded slightly lower effectiveness (MRR@5 = 0.7973/0.6954/0.6340; Recall@5 = 0.7615/0.6343/0.7317), consistent with its more compact architecture. MedBERT placed between the two (MRR@5 = 0.8178/0.6747/0.6381; Recall@5 = 0.8490/0.7921/0.7317). All models exhibited a monotonic decline from training to testing, indicating limited generalization beyond the training distribution. These findings underscore the benefit of domain-specific pre-training for scientific claim re-ranking while highlighting the trade-offs among model capacity, domain specialization, and robustness.

\paragraph{Impact of Batch Size.}
To assess whether optimization stability affects re-ranking accuracy, we fine-tuned SciBERT with batch sizes of 8 and 16, keeping all other settings fixed. Table \ref{tab:batchsize_results} shows that a batch size of 16 yields marginally higher MRR and recall on every split, suggesting that the larger mini-batch provides smoother gradient estimates and slightly better convergence. The gap, however, is small, and test-set performance remains stable, indicating that SciBERT generalizes well under either configuration.

\begin{table}[h]
\centering
\caption{Effect of batch size on SciBERT re-ranking (top-5 candidates).}
\label{tab:batchsize_results}
\begin{tabular}{clccc}
\toprule
\textbf{Batch Size} & \textbf{Dataset} & \textbf{MRR@1} & \textbf{MRR@5} & \textbf{Recall@5} \\
\midrule
\multirow{3}{*}{8}
 & Train & 0.8180 & 0.8300 & 0.8400 \\
 & Dev   & 0.6420 & 0.7000 & 0.7850 \\
 & Test  & 0.6060 & 0.6550 & 0.7250 \\
\midrule
\multirow{3}{*}{16}
 & Train & 0.8216 & \textbf{0.8338} & 0.8490 \\
 & Dev   & 0.6507 & \textbf{0.7070} & 0.7921 \\
 & Test  & 0.6113 & \textbf{0.6607} & 0.7317 \\
\bottomrule
\end{tabular}
\end{table}

\paragraph{Impact of re-ranking depth.}  
Having established the relative strengths of each model, we next investigate how the number of input candidates affects re-ranking performance for our best model (SciBERT). Table~\ref{tab:depth_results} reports MRR@1, MRR@5, and Recall@5 for \(k = 5, 10,\) and 20, highlighting the trade-off between richer candidate pools and the introduction of noise.

\begin{table}[h]
\centering
\caption{Effect of candidate‐set size \(k\) on retrieval metrics with SciBERT.}
\label{tab:depth_results}
\begin{tabular}{llrrrrr}
\toprule
\(k\)  & \textbf{Split} & \textbf{MRR@1 } & \textbf{MRR@5}  & \textbf{MRR@10}  & \textbf{Recall@5} & \textbf{Recall@10}  \\
\midrule
\multirow{3}{*}{5}
       & Train & 0.8216 & 0.8338 & --      & 0.8490    & --         \\
       & Dev   & 0.6507 & 0.7070 & --      & 0.7921    & --         \\
       & Test  & 0.6113 & 0.6607 & --      & 0.7317    & --         \\
\midrule
\multirow{3}{*}{10}
       & Train & 0.8558 & 0.8712 & 0.8716  & 0.8923    & 0.8952     \\
       & Dev   & 0.6529 & \textbf{0.7084} & 0.7149  & 0.7979    & 0.8457     \\
       & Test  & 0.6279 & \textbf{0.6828} & 0.6867  & 0.7656    & 0.7946     \\
\midrule
\multirow{3}{*}{20}
       & Train & 0.8877 & \textbf{0.9031} & 0.9039  & 0.9244    & 0.9303     \\
       & Dev   & 0.6529 & 0.6990 & 0.7064  & 0.7721    & 0.8271     \\
       & Test  & 0.6376 & 0.6711 & 0.6765  & 0.7539    & 0.7826     \\
\bottomrule
\end{tabular}
\end{table}

The results in Table \ref{tab:depth_results} reveal a trade-off between re-ranking depth and generalization. While increasing the candidate set from 5 to 20 leads to steadily higher MRR@1, MRR@5, and Recall@5 on the training split (e.g. MRR@5 rising from 0.8338 at depth 5 to 0.9031 at depth 20), the development and test splits exhibit a peak in performance at the intermediate depth 10. Specifically, performance on the development set improves slightly when moving from depth 5 (MRR@5 = 0.7070) to depth 10 (MRR@5 = 0.7084) but then declines at depth 10 (MRR@5 = 0.6990). A similar pattern can be observed in the test pattern and other metrics such as Recall@5. 

These findings suggest that while larger candidate pools allow the model to memorize more positives during training, they also introduce greater noise and irrelevant documents. This behavior can be explained by the fact that a relatively high proportion of the gold documents were already included in the first 10 results of the previous dual-encoding step. Taking a depth that is higher than necessary can harm re-ranker performance on unseen data. Due to these reasons, a moderate re-ranking depth of 10 was found to best balance training fit and generalization.

\paragraph{E5+HN+All Dual‐Encoder Candidates with SciBERT Re-ranking.}  
Using the optimized dual encoder retrieval stage, the top-5 and top-10 candidates were re-ranked with SciBERT as before. Table~\ref{tab:enhanced_rerank_results} reports the resulting metrics.

\begin{table}[h]
\centering
\caption{Retrieval performance after enhanced dual encoder candidate generation and SciBERT re-ranking for $k=5$ and $k=10$ (batch = 16).}
\label{tab:enhanced_rerank_results}
\begin{tabular}{clrrrrr}
\toprule
\textbf{$k$} & \textbf{Split} & \textbf{MRR@1} & \textbf{MRR@5} & \textbf{MRR@10} & \textbf{Recall@5} & \textbf{Recall@10} \\
\midrule
\multirow{3}{*}{5}
 & Train & 0.8377 & \textbf{0.8522} & --      & 0.8707 & --      \\
 & Dev   & 0.6700 & \textbf{0.7232} & --      & 0.8021 & --      \\
 & Test  & 0.6314 & \textbf{0.6784} & --      & 0.7469 & --      \\
\midrule
\multirow{3}{*}{10}
 & Train & 0.7695 & 0.8115 & 0.8161 & 0.8757 & 0.9091 \\
 & Dev   & 0.6486 & 0.7055 & 0.7117 & 0.7936 & 0.8393 \\
 & Test  & 0.5906 & 0.6559 & 0.6623 & 0.7593 & 0.8064 \\
\bottomrule
\end{tabular}
\end{table}

Finally, we combined the enhanced dual encoder configuration (E5+HN+All) with SciBERT re-ranking at depths k=5 and k=10 (Table \ref{tab:enhanced_rerank_results}). Under this setting, k=5 outperforms k=10 on the development split (MRR@5 = 0.7232 vs.\ 0.7055) and on the test split (MRR@5 = 0.6784 vs.\ 0.6559). The top-5 dense retrieval predictions therefore already cover the majority of relevant documents, and enlarging the pool primarily adds noise, degrading re-ranking accuracy. These results confirm that a shallow re-ranking depth suffices when the upstream retriever is strong.

\subsection{Comparative Model Performance}

The best re-ranking stage was conducted using the SciBERT cross-encoder applied to the top-10 document-candidate set of the base setup dual encoder. This emerged as the best-performing model in the re-ranking experiments. This model achieved a test-set MRR@5 of 0.6828 and an exact-match MRR@1 of 0.6279, markedly outperforming the strongest dual encoder baseline—base setup E5—which recorded MRR@5 = 0.6174 and MRR@1 = 0.5332 as shown in Table \ref{tab:overall-results}. It also reaches the highest Recall@5 with 0.7656. The only measure in which E5+HN+All Fields performs better is Recall@10, likely due to broader document field integration capturing more edge-case matches.

\begin{table}[ht]
\centering
\caption{Overall model comparison on test set.}
\label{tab:overall-results}
\begin{tabular}{lccccc}
\toprule
\textbf{Model} & \textbf{MRR@1} & \textbf{MRR@5} & \textbf{MRR@10} & \textbf{Recall@5} & \textbf{Recall@10} \\
\midrule
Baseline BM25 & 0.4385  & 0.5025 & 0.5097 & 0.5982 & 0.6508  \\
E5+HN+All Fields  & 0.5332 & 0.6174 & 0.6254 & 0.7469 & 0.8064 \\
SciBERT with k=10 & 0.6279 & \textbf{0.6828}& 0.6867 & 0.7656 & 0.7946 \\
\bottomrule
\end{tabular}
\end{table}

To assess the reliability of these improvements, statistical tests were conducted on the per-query performance differences across all test queries. A Wilcoxon signed-rank test on the paired MRR@5 scores yielded a p-value of $p = 1.9 \times 10^{-21}$, confirming that the observed gains are statistically significant. Furthermore, a McNemar's exact test on the top-1 correctness contingency tables yielded 231 queries that were only correct with Sci-Bert and 70 only with E5. This produced a p-value of $p = 3.0 \times 10^{-21}$ which validates that incorporating a SciBert-based re-ranking stage delivers consistent, statistically significant enhancements over the dual encoder baseline.

\begin{figure}[ht]
  \centering
  \def\svgwidth{\textwidth}%
  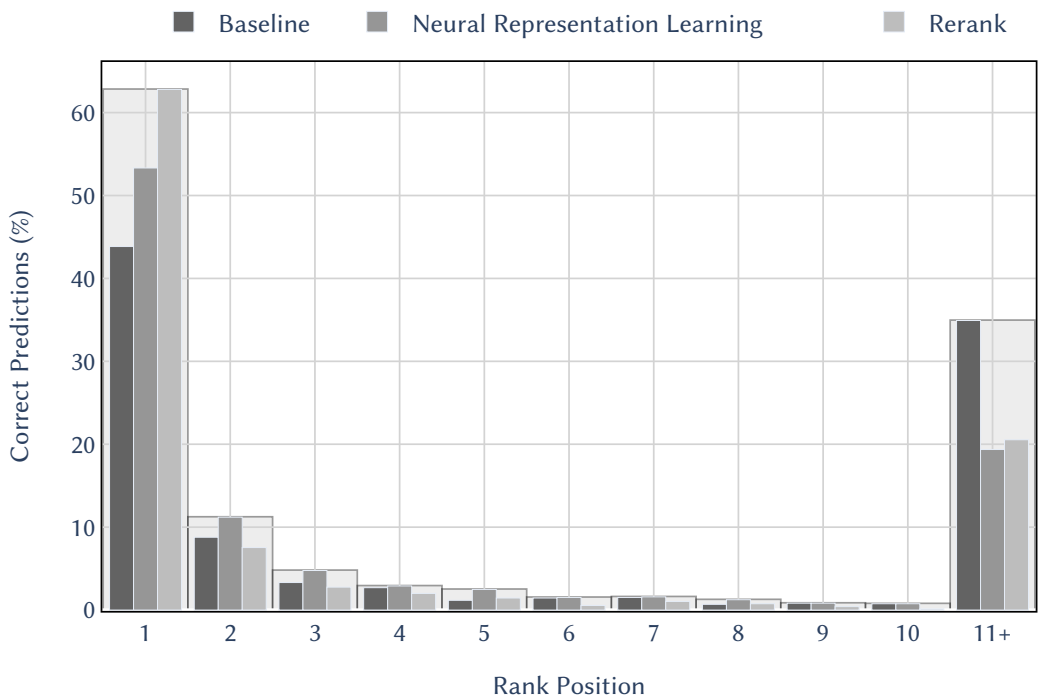
  \caption{Rank Distribution of Correct Predictions (Top-10).}
  \label{fig:rank-distribution}
\end{figure}

Figure \ref{fig:rank-distribution} shows the distribution of the rank position where the correct query was found by the developed information retrieval system. The x-axis represents the rank of the correct prediction within the top 10 results with an additional bar labeled "11+" for cases where the correct document was not found in the top 10, and the y-axis shows the percentage of queries for which the correct article was found at each rank.

Several patterns can be seen in the figure:

\begin{enumerate}
  \item \textbf{Improved Top‐1 Placement:}  
    The dual encoder baseline places the correct document first for approximately 42\% of queries. Neural representation learning raises this rate to about 53\%, and SciBERT re-ranking further increases Top‐1 accuracy to approximately 62\%, confirming its superior fine‐grained discrimination.
  \item \textbf{Steeper Drop‐off Beyond Rank 1:}  
    All systems exhibit a rapid decline after the first position. Both dense retrieval and re-ranking resolve a greater proportion of queries within ranks 2–5 compared to the baseline, demonstrating that semantic models push more correct documents into the early ranks.
  \item \textbf{Reduction in “No‐Hit” Cases:}  
    At rank 11+ (i.e., correct paper not in the top 10), the baseline fails for about 35\% of queries, whereas dense retrieval and re-ranking both reduce no‐hit rates to roughly 20\%.
\end{enumerate}

Note that the SciBERT re-ranker was applied to only the top 10 candidate documents from the dense retrieval. Therefore, it pushed documents that were previously ranked in the first ten positions, thereby improving the Top-1 accuracy while losing accuracy in places 2-10 compared to the dense retrieval. 

Given the presented results, the research questions guiding the present work can now be answered:

\begin{enumerate}
  \item[\textbf{RQ1:}] \textit{Which document metadata is most useful to create embeddings that optimize retrieval performance for the task at hand?}

  Adding the authors field to title + abstract gives the largest standalone gain (MRR@5 rising from 0.6020 to 0.6089), whereas journal metadata contributes the least (MRR@5 to 0.6031) and source falls in between (0.6049). Using all fields together yields a modest overall improvement (MRR@5 to 0.6069). However, using all fields does not reach statistical significance (McNemar’s p=0.2430, Wilcoxon p=0.0715). Thus, while author information appears most informative, title and abstract are the primary drivers of retrieval performance.

  \item[\textbf{RQ2:}] \textit{Does incorporating BM25-mined hard negative training examples for training improve query-to-document retrieval performance?}
  
  Incorporating one BM25-mined hard negative per query into the E5 training significantly boosts test‐set retrieval: MRR@1 rises from 0.5166 to 0.5353 and MRR@5 from 0.6018 to 0.6171, with McNemar’s (p = 0.0159) and Wilcoxon (p = 0.0012) tests confirming these gains are statistically significant. 
  
  \item[\textbf{RQ3:}] \textit{To which degree does re-ranking retrieved results using cross-encoders improve the retrieval performance?}

  Table~\ref{tab:overall-results} and Figure~\ref{fig:rank-distribution} jointly answer this question. When the best dual encoder run (E5+HN+All\,Fields) is enhanced with the SciBERT cross-encoder applied to the top-10 dense candidates, MRR@5 improves from 0.6174 to 0.6828 which is a relative gain of +10.6\%, a Wilcoxon signed-rank test and a McNemar’s exact test confirm that the uplift is highly significant.

  The histogram in \ref{fig:rank-distribution} shows that re-ranking moves roughly an additional 10\% of queries into the first position while also cutting the no-hit rate almost in half, compared to the baseline models.

  The cross-encoder re-ranking models investigated in this paper boost retrieval quality by 0.07 absolute MRR@5 points over an already strong dense retrieval model, with results that are both practically large and statistically significant. 

\end{enumerate}

\section{Conclusion}
\label{sec:conclusion}

This paper presents the experiments conducted by the AIRwaves team for CheckThat! 2025 Task 4b: Scientific Claim Source Retrieval. The approach explored a two-stage IR pipeline leveraging a dual encoder retrieval, followed by neural re-ranking with cross-encoder models, particularly SciBERT, DistilBERT and MedBERT.

Several experiments were conducted with multiple retrieval models, dual encoder configurations (both with and without hard negatives), and varied document representations to address the unique challenges of mapping informal, concise social media posts to their corresponding scholarly sources. Notably, the system achieved a 2nd place ranking out of 31 participating teams on the official leaderboard, underscoring the robustness of the overall approach.

Key findings of the work include that incorporating hard negatives into dual encoder training can improve retrieval performance significantly compared to using only in-batch negative document examples. Furthermore, re-ranking with cross-encoder models like SciBERT consistently boosts mean MRR and overall recall across all splits, highlighting their strength in capturing nuanced query-to-document relationships. Additionally, by experimenting with several different configurations, a practical guideline has been established, demonstrating what might or might not be promising to explore further.

Perspectives for future work that build upon our results are manifold. If the goal is a single-stage retrieval framework due to time and computation constraints that do not allow subsequent re-ranking, there is room for improvement of the phase 1 dual encoder method shown in the present work. For instance, we hypothesize that changes to the hard negative setup could yield further improvements. Our implementation relies on statically mining BM25 hard negative examples per query, where examples deemed ``hard`` by a sparse model might not always be challenging for neural approaches capturing nuanced semantics. A possible improvement might dynamically re-mine hard negative documents using the current model weights after each epoch.

In terms of neural re-ranking, generative methods with large models like MonoT5 were not explored. We encourage future literature to build upon our work by evaluating what retrieval performance is possible when using such models.

The real-world usefulness of the developed approach depends on the specific use case of the system. After re-ranking, roughly 62.8\% posts can be matched to the correct publication right away, as shown by the SciBERT MRR@1. For a zero-shot system that relies on an instantaneous correct answer, this is likely insufficient. In the first 10 suggestions given a query, the correct document out of potentially thousands, will be contained roughly 79.5\% of the time. If manual postprocessing of a few documents is an option, the developed system might therefore be tremendously helpful in narrowing down the search space.

Overall, the study demonstrates that hybrid retrieval strategies, combined with careful sampling and neural re-ranking, are powerful tools for linking social media claims to relevant scientific literature.

\section*{Declaration on Generative AI}

During preparation the authors followed the CEUR‐WS activity taxonomy\footnote{\url{https://ceur-ws.org/genai-tax.html}} and used Grammarly for grammar and spelling checks, and ChatGPT (GPT-4o) for writing-style improvement, citation management, and formatting assistance. After using these tools, the authors reviewed and edited the content as needed and take full responsibility for the publication’s content.

\bibliography{sample-ceur}

\appendix

\appendix



\end{document}